\documentclass[twocolumn,twoside,slac_two]{revtex4}
\usepackage{graphicx}
\usepackage{fancyhdr}
\begin{document}

\title{Recent studies of Charmonium Decays at CLEO}

%

\author{H. Muramatsu}
\affiliation{Department of Physics and Astronomy, University of Rochester, Rochester, New York 14627, USA}

%
%

\begin{abstract}
Recent results on Charmonium decays are reviewed which includes two-,
three- and four-body decays of $\chi_{cJ}$ states, observations of
Y(4260) through $\pi\pi J/\psi$ transitions, precise measurements of
$M(D^0)$, $M(\eta)$ as well as $\mathcal{B}(\eta\to X)$.
\end{abstract}

\maketitle

\thispagestyle{fancy}


\section{Introduction}

Decays of a bound state of a quark and its anti-quark, quarkonium, 
provide an excellent laboratory for studying QCD. Particularly,
heavy quarkonia such as charmonium states are less relativistic, thus
play a special role in probing strong interactions.

CLEO recently has accumulated data taken at the $\psi(2S)$ resonance, providing
a total of 27M $\psi(2S)$ decays. With the combination of this 
large statistical
sample and the excellent CLEO detector, we will explore an unprecedented world
of charmonia. While many analyses are currently being carried out,
in this note we present recent results on multi-body $\chi_{cJ}$ decays
which employed the pre-existing 3M $\psi(2S)$ sample.

We also present recent studies on decays of one of the exotic states,
Y(4260), as well as precision measurement on $M(D^0)$ that has an implication
on properties of X(3872).

Finally, based on the full sample of $\psi(2S)$ data, we have results
on properties of one of the light mesons, $\eta$.

\section{Factory of $\chi_c(1^3P_J)$ states}

 $\chi_c(1^3P_J)$ states, which have one unit of orbital angular
momentum and total spin of J=0, 1, or 2, cannot be produced
directly from $e^+e^-$ collisions. They can be reached from
$\psi(2S)$ through radiative (electric dipole) transitions.
Since $\mathcal{B}(\psi(2S)\to\gamma\chi_{cJ})=(9.3\pm0.4,
8.8\pm0.4$, and $8.1\pm0.4)\times 10^{-2}$ for J=0, 1, and 2
respectively \cite{pdg06}, 27M $\psi(2S)$ decays of the new data
provides $\sim$2M decays of each spin state of $\chi_{cJ}$ which should
give us a greater understanding of the decay mechanisms of the
$\chi_{cJ}$ mesons.

In this section, we present recent results of studies of $\chi_{cJ}$
decays based on 3M $\psi(2S)$ decays which should serve as
the foundation for the future precision measurements by employing
the full data sample of 27M of $\psi(2S)$ decays.

\subsection{Two-body decay}
\begin{figure}[h]
\centering
\includegraphics[width=80mm]{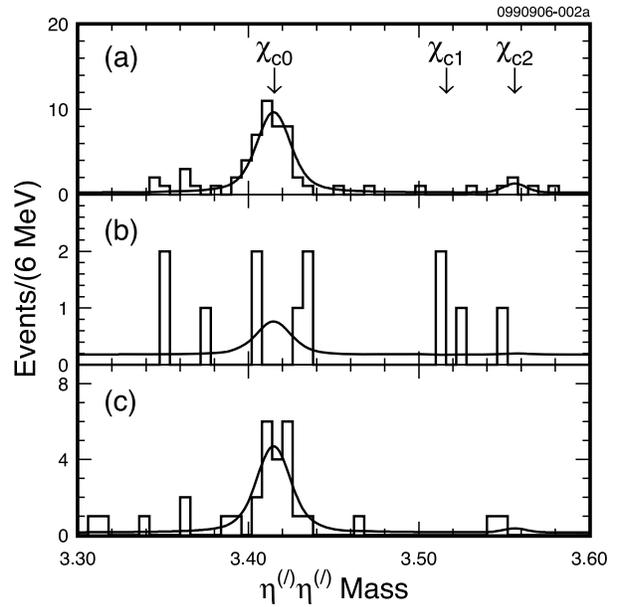}
\caption{Invariant masses of $\eta\eta$ (a), $\eta'\eta$ (b),
and $\eta'\eta'$ (c)} \label{fig:etaetamass}
\end{figure}

We present results on $\chi_{cJ}$ decay into combinations of
$\eta$ and $\eta'$ mesons. Figure~\ref{fig:etaetamass} shows
invariant masses of combinations of $\eta$ and $\eta'$.
No $\chi_{c1}$ is seen as expected from conservation of spin-parity.

We measured $\mathcal{B}(\chi_{c0}\to\eta\eta)$ to be
$(0.31\pm0.05\pm0.04\pm0.02)\%$~\cite{etaeta} where the first uncertainty
is statistical, the second is systematic, and the third is systematic due to
the uncertainty in $\mathcal{B}(\psi(2S)\to\gamma\chi_{cJ})$.
This is slightly higher, but
consistent with, the two previously published measurements.
The BES Collaboration measured this branching ratio to be 
$(0.194\pm0.085\pm0.059)\%$~\cite{beseta} and
the E-835 Collaboration~\cite{e823eta} had $(0.198\pm0.068\pm0.037)\%$.
We also measured $\mathcal{B}(\chi_{c0}\to\eta'\eta')$ to be
$(0.17\pm0.04\pm0.02\pm0.01)\%$ for the first time.
We set upper limits for 
$\mathcal{B}(\chi_{c0}\to\eta\eta')<0.05\%$ ,
$\mathcal{B}(\chi_{c2}\to\eta\eta)<0.047\%$,
$\mathcal{B}(\chi_{c2}\to\eta\eta')<0.023\%$, and
$\mathcal{B}(\chi_{c2}\to\eta'\eta')<0.031\%$
at $90\%$ confidence level.

Our result can be compared to predictions based on the model of
Qiang Zhao~\cite{zhao}. He translates these decay rates into
a QCD parameter, $r$, which is the ratio of doubly- to  singly-OZI
suppressed decay diagrams. In his model, our results indicate
that the singly-OZI suppressed diagram dominates in these decays.

\subsection{Three-body decay}

We have also looked at three-body decays of $\chi_{cJ}$ states
(one neutral and 2 charged hadrons)~\cite{dalitz}. They are
$\pi^+\pi^-\eta$, $K^+K^-\eta$, $p\bar{p}\eta$, $\pi^+\pi^-\eta'$,
$K^+K^-\pi^0$, $p\bar{p}\pi^0$, $\pi^+K^-K^0_S$, and
$K^+\bar{p}\Lambda$.
Measured branching fractions are summarized in Table~\ref{tab:3btab}.
Again, our results are consistent with the results from BES
Collaboration~\cite{bes3body}, with better precision.

\begin{figure}[h]
\centering
\includegraphics[width=40mm]{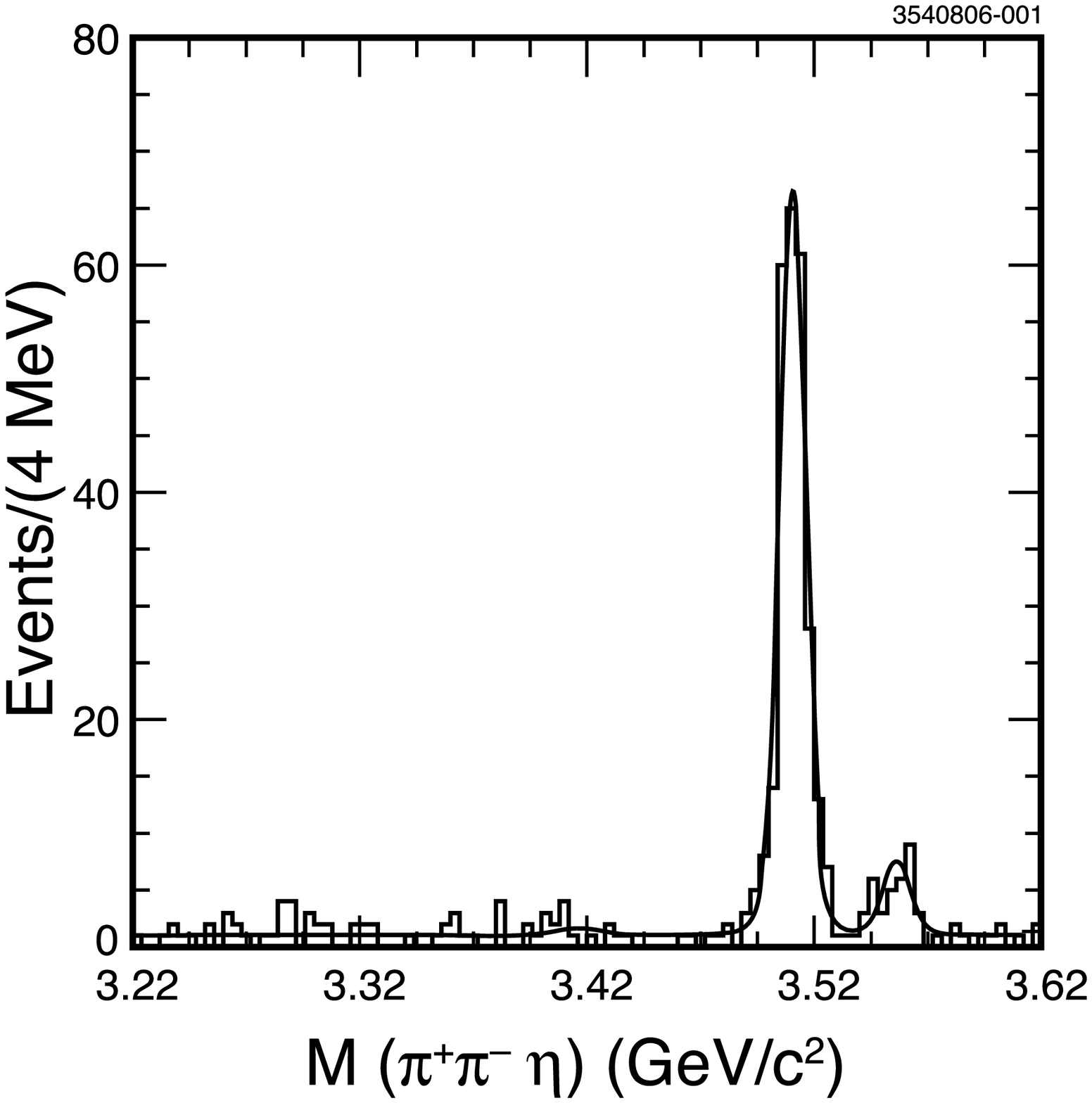}
\includegraphics[width=40mm]{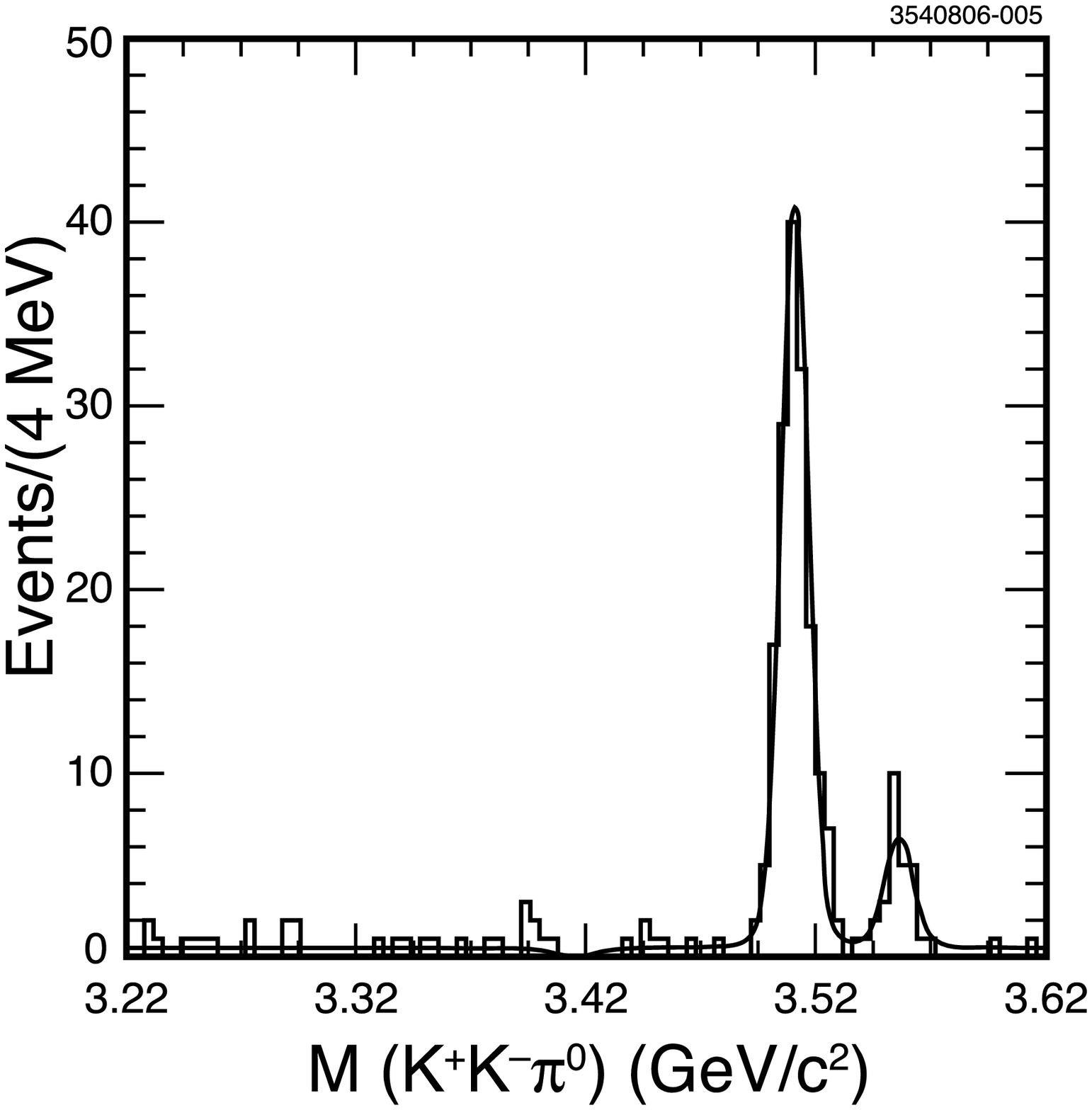}
\includegraphics[width=40mm]{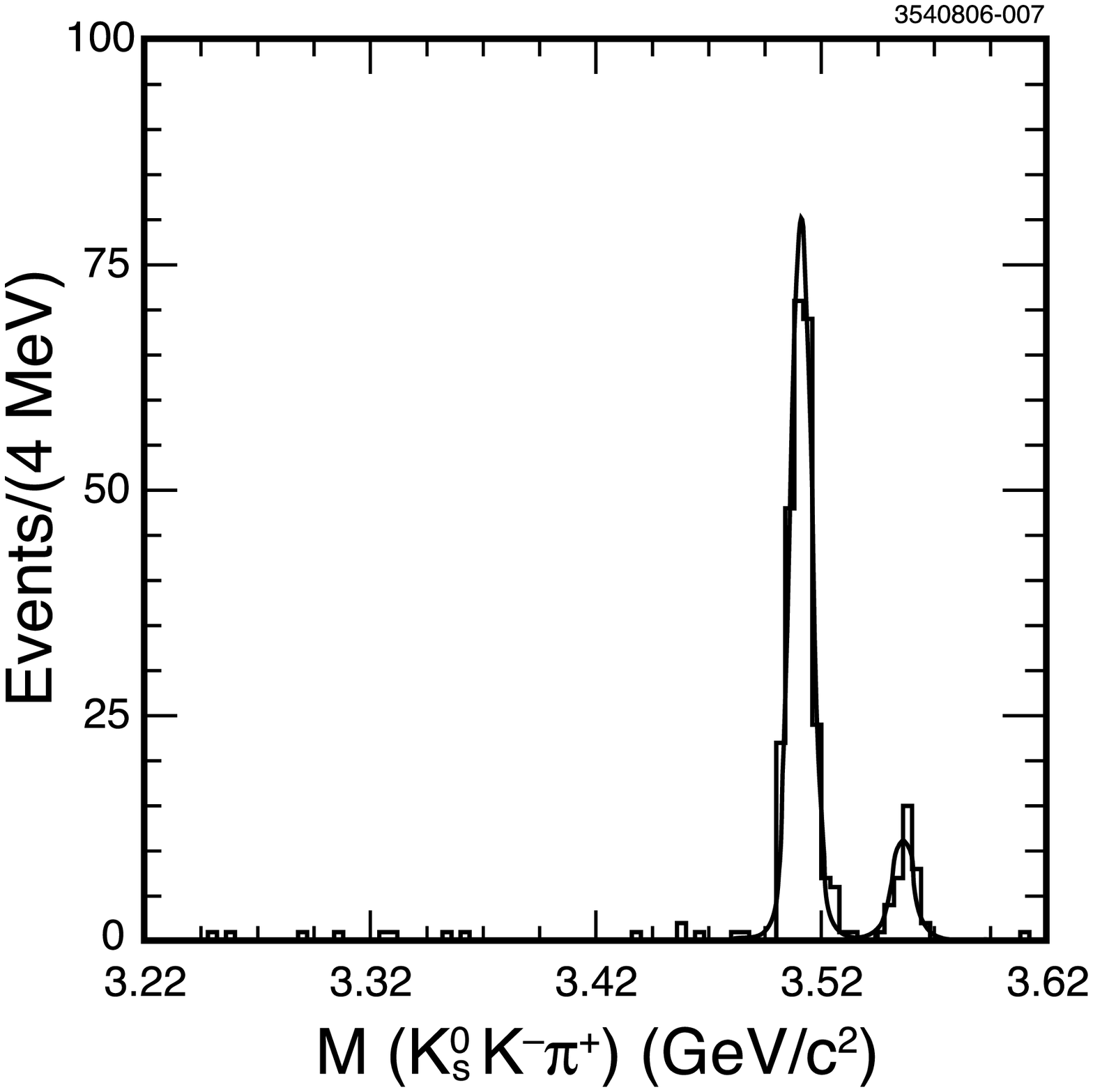}
\caption{Invariant masses of $\pi^+\pi^-\eta$ (top-left), 
$K^+K^-\pi^0$ (top-right),
and $K^0_SK^-\pi^+$ (bottom)} \label{fig:threebody}
\end{figure}

In three of the above modes we looked for, $\pi^+\pi^-\eta$,
$K^+K^-\pi^0$, and $\pi^+K^-K^0_S$, we observed significant
signals of $\chi_{c1}$ decays which are shown in Figure~\ref{fig:threebody}.

\begin{table*}[t]
\begin{center}
\caption{Branching fractions in 
units of 10$^{-3}$.  Uncertainties are statistical,
systematic due to detector effects plus analysis methods,
and a separate systematic due to uncertainties in the $\psi(2S)$ branching
fractions.  Limits are at the 90\% confidence level.}
\begin{tabular}{|l|c|c|c|}
\hline \hline
Mode &$\chi_{c0}$ & $\chi_{c1}$ & $\chi_{c2}$ \\
\hline
$\pi^+\pi^-\eta $         & $<0.21$  
                          & $5.0 \pm 0.3 \pm 0.4 \pm 0.3$ 
                          & $0.49\pm 0.12\pm 0.05\pm 0.03$ \\
$K^+K^-\eta$              & $<0.24$  
                          & $0.34\pm 0.10\pm 0.03\pm 0.02$ 
                          & $<0.33$                        \\
$p\bar{p}\eta$            & $0.39\pm 0.11\pm 0.04\pm 0.02$  
                          & $<0.16$ 
                          & $0.19\pm 0.07\pm 0.02\pm 0.01$ \\
$\pi^+\pi^-\eta^{\prime}$ & $<0.38$  
                          & $2.4 \pm 0.4  \pm0.2 \pm 0.2$     
                          & $0.51\pm 0.18 \pm0.05\pm 0.03$ \\
$K^+K^-\pi^0$             & $<0.06$  
                          & $1.95\pm 0.16\pm 0.18\pm 0.14$ 
                          & $0.31\pm 0.07\pm 0.03\pm 0.02$ \\
$p\bar{p}\pi^0$           & $0.59\pm 0.10\pm 0.07\pm 0.03$ 
                          & $0.12\pm 0.05\pm 0.01\pm 0.01$ 
                          & $0.44\pm 0.08\pm 0.04\pm 0.03$ \\
$\pi^+K^-\overline{K}^0$  & $<0.10$  
                          & $8.1 \pm 0.6 \pm 0.6 \pm 0.5$ 
                          & $1.3 \pm 0.2 \pm 0.1 \pm 0.1$  \\
$K^+ \bar{p}\Lambda$      & $1.07\pm 0.17\pm 0.10\pm 0.06$ 
                          & $0.33\pm 0.09\pm 0.03\pm 0.02$ 
                          & $0.85\pm 0.14\pm 0.08\pm 0.06$ \\ 
\hline \hline
\end{tabular}
\label{tab:3btab}
\end{center}
\end{table*}

\begin{figure}[h]
\centering
\includegraphics[width=80mm]{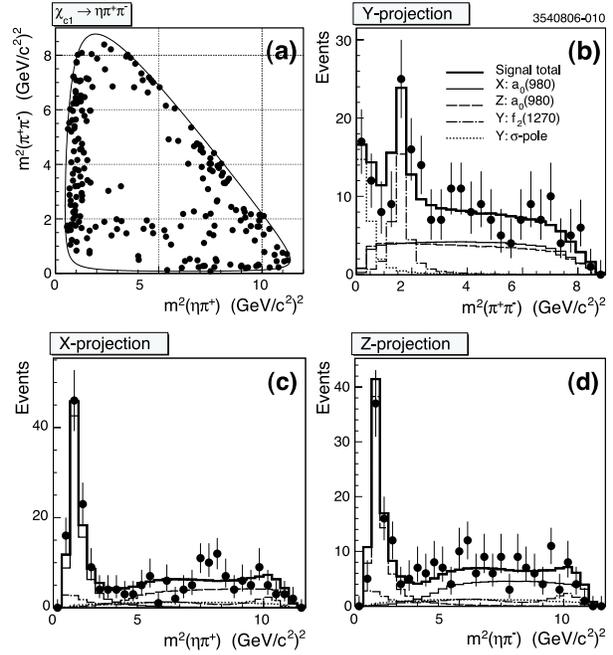}
\caption{Dalitz plots for  $\chi_{c1} \to \eta\pi^+\pi^-$.} \label{fig:ppeta-dal}
\end{figure}

We performed Dalitz plot analyses based on these 3 modes in which
we neglected any possible interference effects between resonances
and polarization of $\chi_{c1}$. We estimated there could be
$\sim 20(15)\%$ variations in fit fractions for the $\pi\pi\eta$ ($KK\pi$)
mode due to such a simplified model.
Figure~\ref{fig:ppeta-dal} shows the Dalitz plot for $\pi^+\pi^-\eta$
and Table~\ref{tab:pipietadalitz} shows its resultant fit fractions for
each source. It is interesting to note that our data demand
a relatively large yield of a $\sigma$ pole.

\begin{table}
\caption{Fit results for $\chi_{c1} \to \eta\pi^+\pi^-$ Dalitz plot analysis.  The uncertainties
         are statistical and systematic.
         Allowing for interference among the resonances changes the fit fractions
         by as much as 20\% in absolute terms as discussed in the text.
        }
\begin{tabular}{l|c}
\hline \hline
Mode                   & Fit Fraction (\%)  \\ \hline
$a_0(980)^\pm \pi^\mp$ & $75.1\pm3.5\pm4.3$ \\
$f_2(1270) \eta$       & $14.4\pm3.1\pm1.9$ \\    
$\sigma \eta$          & $10.5\pm2.4\pm1.2$ \\
\hline \hline
\end{tabular}
\label{tab:pipietadalitz} 
\end{table}

\begin{figure}[h]
\centering
\includegraphics[width=80mm]{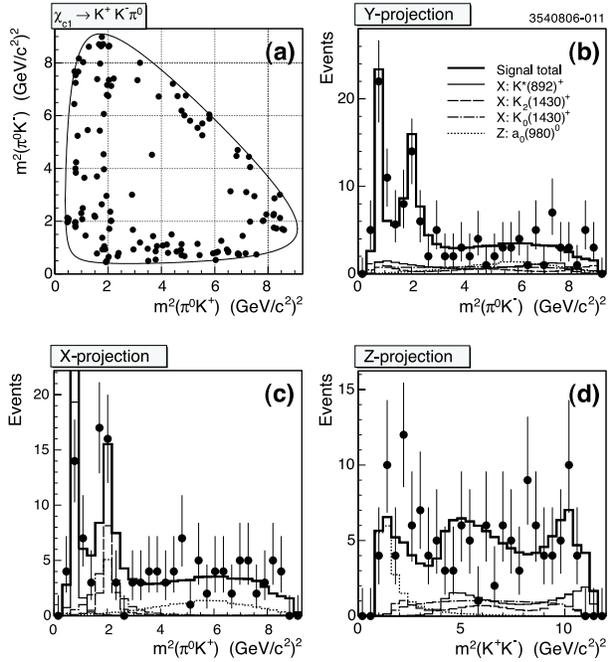}
\caption{Dalitz plots for $\chi_{c1} \to K^+ K^- \pi^0$.} \label{fig:kkp0-dal}
\end{figure}

\begin{figure}[h]
\centering
\includegraphics[width=80mm]{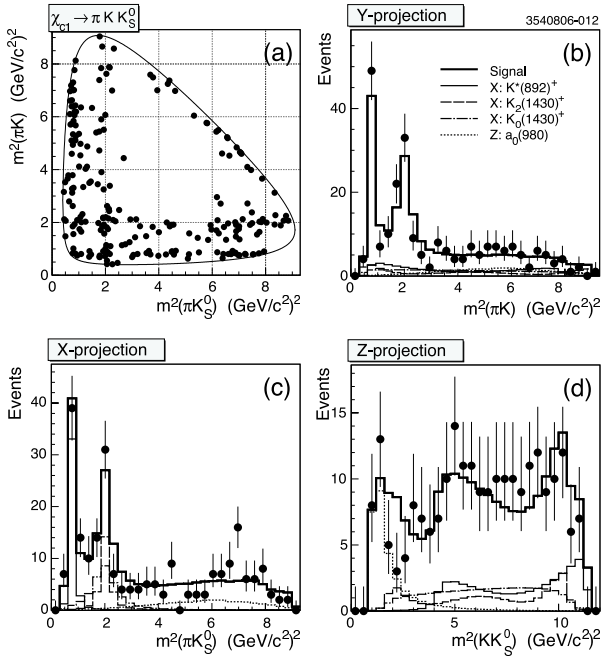}
\caption{Dalitz plots for  $\chi_{c1} \to \pi K K^0_S$.} \label{fig:kskp-dal}
\end{figure}

As for the $KK\pi$ mode, we performed simultaneous fits between
$\chi_{c1}\to K^+K^-\pi^0$ and $\chi_{c1}\to K_SK\pi$ by taking
advantage of isospin symmetry. Dalitz plots for these modes are
shown in Figures~\ref{fig:kkp0-dal} and \ref{fig:kskp-dal}.
Table~\ref{tab:KKpiDalitz} shows their resultant
fit fractions.


\begin{table}[!htb]
\caption{Results of the combined fits to the 
$\chi_{c1} \to K^+ K^- \pi^0$ and $\chi_{c1} \to \pi K K^0_S$ Dalitz plots.
         Allowing for interference among the resonances changes the fit fractions
         by as much as 15\% in absolute terms as discussed in the text.}
\begin{tabular}{l|c}
\hline \hline
Mode                   & Fit Fraction (\%)   \\ \hline
$K^*(892)K$            & $31.4\pm2.2\pm1.7$  \\
$K^*_0(1430)K$         & $30.4\pm3.5\pm3.7$ \\
$K^*_2(1430)K$         & $23.1\pm3.4\pm7.1$  \\
$a_0(980)\pi$          & $15.1\pm2.7\pm1.5$  \\
\hline \hline
\end{tabular}
\label{tab:KKpiDalitz}
\end{table}

\subsection{Four-body decay}
We present a preliminary result on four-body decay of
$\chi_{cJ}$ states in which we reconstructed
$h^+h^-\pi^0\pi^0$, where $h=\pi$, $K$, $p$; $K^+K^-\eta\pi^0$; and
$K^{\pm}\pi^{\mp}K^0_S\pi^0$. Results of this kind of study, many-body
decays of $\chi_{cJ}$ states, should help to build a comprehensive
understanding about the P-wave dynamics.

\begin{figure}[h]
\centering
\includegraphics[width=80mm]{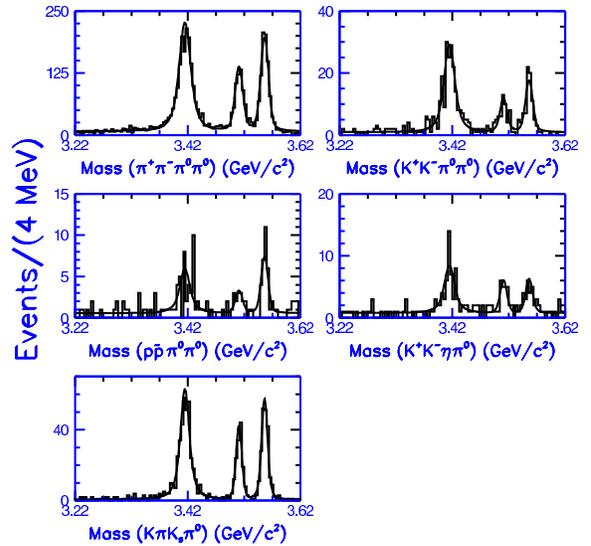}
\caption{Preliminary result on 4-body decays of $\chi_{cJ}$.
Invariant masses of various combinations of hadrons
are shown.} \label{fig:mass4body}
\end{figure}

Clean signals were seen in all modes except 
$\chi_{c1}\to p\bar{p}\pi^0\pi^0$
for the first time as can be seen in
Figure~\ref{fig:mass4body}. 
Many resonant substructures were also seen
for which we only considered significant ones ($\pi^+\pi^-\pi^0\pi^0$
and $K^{\pm}\pi^{\mp}K^0_S\pi^0$). The results are summarized in
Table~\ref{Table:tableofBFsall}. The measured branching fraction of
$\chi_{cJ}\to\rho^{\pm}\pi^{\mp}\pi^0$ is consistent with that of
$\chi_{cJ}\to\rho^0\pi^+\pi^-$ as expected from isospin symmetry.
Similar isospin symmetry is also seen in Table~\ref{table:table_br_isospin}
where the partial width of $\chi_c\to K^{*0}K^0\pi^0$ and that of
$\chi_c\to K^{*\pm}K^{\mp}K^0$ are expected to be equal.
Table~\ref{table:table_br_isospin} also shows another good agreement with
the isospin expectation of $\mathcal{B}(\chi_c\to K^{*0}K^0\pi^0)/$
$\mathcal{B}(\chi_c\to K^{*0}K^{\pm}\pi^{\mp})$ = 0.5 and
$\mathcal{B}(\chi_c\to K^{*0}K^0\pi^0)/$
$\mathcal{B}(\chi_c\to K^{*\pm}\pi^{\mp}K^0)$ = 0.5.

\begin{table}[htb]
\caption{Branching fractions and combined error measurements for the isospin
related $K^{*}K\pi$ intermediate modes are listed.}
\begin{tabular}{|l|c|c|c|}
\hline
Mode  &{ $\chi_{c0}$ }& {$\chi_{c1}$} &{ $\chi_{c2}$} \\
\hline
&B.F. (\%) & B.F. (\%) & B.F. (\%)\\
\hline
$K^{*0}K^0\pi^0$ &0.56$\pm$0.15 &0.38$\pm$0.11 &0.59$\pm$0.14 \\
$K^{*0}K^{\pm}\pi^{\mp}$ &- &-  &0.90$\pm$0.25 \\
$K^{*\pm}K^{\mp}\pi^0$ &0.74$\pm$0.18 &-  &0.57$\pm$0.13 \\
$K^{*\pm}\pi^{\mp}K^0$ &0.96$\pm$0.25 &- &0.90$\pm$0.25\\
\hline
\end{tabular}
\label{table:table_br_isospin}
\end{table}

\begin{table*}[htb]
 \caption{Branching fractions (B.F.) with statistical and systematic
 uncertainties are shown. The symbol ``$\times$'' indicates product of
 B.F.'s. The third error in each case is
 due to the $\psi(2S) \to \gamma \chi_{c} $ branching fractions. Upper limits
 shown are at 90\% C.L and include all the systematic errors.
 The measurements of the three-hadron final states are inclusive branching
 fractions, and do not represent the amplitudes for the three-body
 non-resonant decays.}
 \begin{tabular}{|l|c|c|c|}
 \hline
 Mode  & { $\chi_{c0}$ }& {$\chi_{c1}$} & { $\chi_{c2}$} \\
 \hline
 & B.F.($\%$) & B.F.($\%$) & B.F.($\%$) \\
 \hline
 $\pi^+\pi^-\pi^0\pi^0$
 &    3.54$\pm$0.10$\pm$0.43$\pm$0.18
 &    1.28$\pm$0.06$\pm$0.16$\pm$0.08
 &    1.87$\pm$0.07$\pm$0.23$\pm$0.13 \\
 &
 &
 & \\
 $\rho^+\pi^-\pi^0$
 &1.48$\pm$0.13$\pm$0.18$\pm$0.08
 &0.78$\pm$0.09$\pm$0.09$\pm$0.05
 &1.12$\pm$0.08$\pm$0.14$\pm$0.08 \\
 $\rho^-\pi^+\pi^0$
 &1.56$\pm$0.13$\pm$0.19$\pm$0.08
 &0.78$\pm$0.09$\pm$0.09$\pm$0.05
 &1.11$\pm$0.09$\pm$0.13$\pm$0.08 \\
 \hline
 $K^+K^-\pi^0\pi^0$
 &    0.59$\pm$0.05$\pm$0.08$\pm$0.03
 &    0.12$\pm$0.02$\pm$0.02$\pm$0.01
 &    0.21$\pm$0.03$\pm$0.03$\pm$0.01 \\
 \hline
 $p\bar{p}\pi^0\pi^0$
 &    0.11$\pm$0.02$\pm$0.02$\pm$0.01
 &    $< 0.05 $
 &    0.08$\pm$0.02$\pm$0.01$\pm$0.01 \\
 \hline
 $K^+K^-\eta\pi^0$
 &    0.32$\pm$0.05$\pm$0.05$\pm$0.02
 &    0.12$\pm$0.03$\pm$0.02$\pm$0.01
 &    0.13$\pm$0.04$\pm$0.02$\pm$0.01 \\
 \hline
 $K^{\pm}\pi^{\mp}K^0\pi^0$
 &    2.64$\pm$0.15$\pm$0.31$\pm$0.14
 &    0.92$\pm$0.09$\pm$0.11$\pm$0.06
 &    1.41$\pm$0.10$\pm$0.16$\pm$0.10 \\
 &
 &
 & \\
 $K^{*0}K^0\pi^0$ $\times$ $K^{*0}\to K^{\pm}\pi^{\mp}$
  &0.37$\pm$0.09$\pm$0.04$\pm$0.02
 &0.25$\pm$0.06$\pm$0.03$\pm$0.02
 &0.39$\pm$0.07$\pm$0.05$\pm$0.03 \\
 $K^{*0}K^{\pm}\pi^{\mp}$ $\times$ $K^{*0}\to K^0\pi^0$
 &
 &
 &0.30$\pm$0.07$\pm$0.04$\pm$0.02 \\
 $K^{*\pm}K^{\mp}\pi^0$ $\times$ $K^{*\pm}\to \pi^{\pm}K^0$
 &0.49$\pm$0.10$\pm$0.06$\pm$0.03
 &
 &0.38$\pm$0.07$\pm$0.04$\pm$0.03 \\
 $K^{*\pm}\pi^{\mp}K^0$ $\times$ $K^{*\pm}\to K^{\pm}\pi^0$
 &0.32$\pm$0.07$\pm$0.04$\pm$0.02
 &
 &0.30$\pm$0.07$\pm$0.04$\pm$0.02 \\
 $\rho^{\pm}K^{\mp}K^0$
 &1.28$\pm$0.16$\pm$0.15$\pm$0.07
  &0.54$\pm$0.11$\pm$0.06$\pm$0.03
  &0.42$\pm$0.11$\pm$0.05$\pm$0.03 \\
 \hline
 \end{tabular}
 \label{Table:tableofBFsall}
 \end{table*}

\section{Charmonium-like states above $\mbox{D}\bar{\mbox{D}}$}

\subsection{Y(4260)}

\begin{figure}[h]
\centering
\includegraphics[width=80mm]{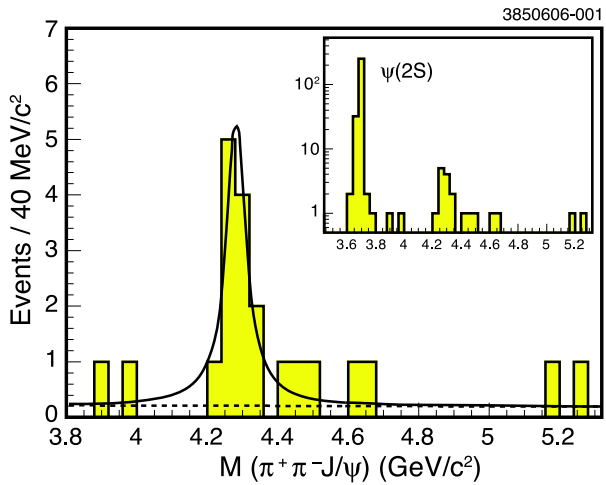}
\caption{Distribution of invariant mass of 
$\pi^+\pi^-J/\psi$ produced in ISR.} \label{fig:y4260isr}
\end{figure}

Y(4260) was first discovered by the BaBar Collaboration via the reaction of
$e^+e^-\to\gamma Y(4260) \to \gamma \pi^+\pi^-J/\psi$,
$J/\psi\to\ell^+\ell^-$~\cite{y4260ba}. Through the same production mechanism
of initial state radiation, we also confirmed their observation based on
data taken around the $\Upsilon(nS)$ resonances, where $n$ is 1, 2, 3, and, 
4~\cite{y4260isr}. The invariant mass of $\pi^+\pi^-J/\psi$ based on such ISR
production is shown in Figure~\ref{fig:y4260isr}.

We further confirmed BaBar's observation in data taken directly at 
$\sqrt{s}=4260$~MeV~\cite{y4260sigma} at 11 $\sigma$ significance.
We also observed $Y(4260)\to\pi^0\pi^0J/\psi$ for the first time at
significance of 5.1 $\sigma$ and had the first evidence for
$Y(4260)\to K^+K^- J/\psi$ ($3.7\sigma$).
We measured $e^+e^-$ cross sections as
$\sigma(\pi^+\pi^-J/\psi)=(58^{+12}_{-10}\pm4)$~pb,
$\sigma(\pi^0\pi^0J/\psi)=(23^{+12}_{-8}\pm1)$~pb, and
$\sigma(K^+K^-J/\psi)=(9^{+9}_{-5}\pm1)$~pb.
Using scan data taken between $\sqrt{s}$ of 3970~MeV and 4260~MeV,
we also searched in 12 additional modes (transitions down to
$\psi(2S)$, $\chi_{cJ}$, and $J/\psi$). No evidence of
strong signals were seen for which we set upper limits at $90\%$ confidence
level. The results are summarized in Table~\ref{tab:y4260}.

\begin{table*}[htp]
\setlength{\tabcolsep}{0.4pc}
\catcode`?=\active \def?{\kern\digitwidth}
\caption{
For each mode $e^+e^- \to X$, for three center-of-mass regions:
the detection efficiency, $\epsilon$; 
the number of signal [background] events in data, 
$N_{\mathrm{s}}$ 
[$N_{\mathrm{b}}$];
the cross-section $\sigma(e^+e^- \to X)$; 
and the branching fraction of $\psi(4040)$ or $\psi(4160)$ to $X$,
${\cal B}$. Upper limits are at 90\%~CL. '--' indicates that the
channel is kinematically or experimentally inaccessible.
}
\label{tab:y4260}
\footnotesize
\begin{center}
\begin{tabular}{c|ccccc|ccccc|cccc}
\hline
\hline
 & \multicolumn{5}{c|}{$\sqrt s = 3970-4060$~MeV} 
 & \multicolumn{5}{c|}{$\sqrt s = 4120-4200$~MeV} 
 & \multicolumn{4}{c}{$\sqrt s = 4260$~MeV} 
 \\ 
Channel 
& $\epsilon$ & $N_{\mathrm{s}}$ & $N_{\mathrm{b}}$ & $\sigma$ & ${\cal B}$ 
& $\epsilon$ & $N_{\mathrm{s}}$ & $N_{\mathrm{b}}$ & $\sigma$ & ${\cal B}$ 
& $\epsilon$ & $N_{\mathrm{s}}$ & $N_{\mathrm{b}}$ & $\sigma$ 
 \\ 
        
& ($\%$)       &                 &                       & (pb)   &     $(10^{-3})$ 
& ($\%$)       &                 &                       & (pb)   &     $(10^{-3})$ 
& ($\%$)       &                 &                       & (pb)        
 \\ 
\hline
 $ \pi^+\pi^-J/\psi $ 
       & 37 &  12 &  3.7 &  $    9^{+5}_{-4}$$\pm 2 $ &  $ < 4 $ 
       & 38 &  13 &  3.7 &  $    8^{+4}_{-3}$$\pm 2 $ &  $ < 4 $ 
       & 38 &  37 &  2.4 &  $   58^{+12}_{-10}$$\pm 4 $ 

 \\ 
 $ \pi^0\pi^0J/\psi $ 
       & 20 &   1 &  1.9 &  $ < 8 $  &  $ < 2 $ 
       & 21 &   5 &  0.9 &  $    6^{+5}_{-3}$$\pm 1 $ &  $ < 3 $ 
       & 22 &   8 &  0.3 &  $   23^{+12}_{-8}$$\pm 1 $ 

 \\ 
 $ K^+K^-J/\psi $ 
       & \multicolumn{5}{c|}{--} 
       & 7 &   1 &  0.07 &  $ < 20 $  &  $ < 5 $ 
       & 21 &   3 &  0.07 &  $    9^{+9}_{-5}$$\pm 1 $ 

 \\ 
 $ \eta J/\psi $ 
       & 19 &  12 &  9.5 &  $ < 29 $  &  $ < 7 $ 
       & 16 &  15 &  8.8 &  $ < 34 $  &  $ < 8 $ 
       & 16 &   5 &  2.7 &  $ < 32 $  

 \\ 
 $ \pi^0J/\psi $ 
       & 23 &   2 &         &  $ < 10 $  &  $ < 2 $ 
       & 22 &   1 &         &  $ < 6 $  &  $ < 1 $ 
       & 22 &   1 &         &  $ < 12 $

 \\ 
 $ \eta'J/\psi $ 
       & \multicolumn{5}{c|}{--} 
       & 11 &   4 &  2.5 &  $ < 23 $  &  $ < 5 $ 
       & 11 &   0 &  1.5 &  $ < 19 $  

 \\ 
 $ \pi^+\pi^-\pi^0J/\psi $ 
       & 21 &   1 &         &  $ < 8 $  &  $ < 2 $ 
       & 21 &   0 &         &  $ < 4 $  &  $ < 1 $ 
       & 22 &   0 &         &  $ < 7 $  

 \\ 
 $ \eta\eta J/\psi $ 
       & \multicolumn{5}{c|}{--} 
       & \multicolumn{5}{c|}{--} 
       & 6 &   1 &         &  $ < 44 $  

 \\ 
 $ \pi^+\pi^-\psi(2S)$ 
       & \multicolumn{5}{c|}{--} 
       & 12 &   0 &         &  $ < 15 $  &  $ < 4 $ 
       & 19 &   0 &         &  $ < 20 $  

 \\ 
 $ \eta\psi(2S) $ 
       & \multicolumn{5}{c|}{--} 
       & \multicolumn{5}{c|}{--} 
       & 15 &   0 &         &  $ < 25 $  

 \\ 
 $ \omega\chi_{c0} $ 
       & \multicolumn{5}{c|}{--} 
       & \multicolumn{5}{c|}{--} 
       & 9 &  11 & 11.5 &  $ < 234 $  

 \\ 
 $ \gamma\chi_{c1} $ 
       & 26 &   9 &  8.1 &  $ < 50 $  &  $ < 11 $ 
       & 26 &  11 &  8.7 &  $ < 45 $  &  $ < 10 $ 
       & 26 &   1 &  3.3 &  $ < 30 $  

 \\ 
 $ \gamma\chi_{c2} $ 
       & 25 &   6 &  8.0 &  $ < 76 $  &  $ < 17 $ 
       & 26 &  10 &  8.6 &  $ < 79 $  &  $ < 18 $ 
       & 27 &   4 &  3.3 &  $ < 90 $  

 \\ 
 $ \pi^+\pi^-\pi^0\chi_{c1}$ 
       & 6 &   0 &         &  $ < 47 $  &  $ < 11 $ 
       & 8 &   0 &         &  $ < 26 $  &  $ < 6 $ 
       & 9 &   0 &         &  $ < 46 $  

 \\ 
 $ \pi^+\pi^-\pi^0\chi_{c2} $ 
       & 4 &   0 &         &  $ < 141 $  &  $ < 32 $ 
       & 8 &   0 &         &  $ < 56 $  &  $ < 13 $ 
       & 9 &   0 &         &  $ < 96 $  

 \\ 
 $ \pi^+\pi^-\phi $ 
       & 17 &  26 &  3.0 &  $ < 12 $  &  $ < 3 $ 
       & 17 &  17 &  6.0 &  $ < 5 $  &  $ < 1 $ 
       & 18 &   7 &  5.5 &  $ < 5 $  

 \\ 
\hline
\hline
\end{tabular} 
\end{center}
\end{table*}

The observation of $Y(4260)\to\pi^0\pi^0J/\psi$ is inconsistent with
the $\chi_{cJ}\rho^0$ molecular model~\cite{chirho}.
Our observation of $\pi^0\pi^0J/\psi$ rate being about half of
$\pi^+\pi^-J/\psi$ rate disagrees with the prediction of the
baryonium model~\cite{baryon}. Evidence for the $K^+K^-J/\psi$ signal
is not compatible with these two models either.
Table~\ref{tab:y4260} also shows that Y(4160) behaves very differently
compared to other charmonium states above $D\bar{D}$ threshold such as
$\psi(4040)$ and $\psi(4160)$ for which we set upper limits in terms
of cross section ($\sigma(e^+e^-\to X)$) and branching fractions.

\subsection{X(3872) and Mass of neutral D meson}

Since X(3872) was discovered by Belle Collaboration~\cite{x3872bell} and
subsequently confirmed by other experiments~(\cite{x3872cdf},\cite{x3872d0},\cite{x3872baba}), many theoretical models have been proposed. Perhaps the most
provocative theoretical suggestion is that X(3872) is a loosely bound state of
$D^0$ and $\bar{D^{*0}}$ mesons~\cite{ddbound}. This idea arises mainly because
$M(D^0) + M(D^{*0}) - M(X(3872))$ is very small. Using the average value
of $M(D^0)$ of 2006 Particle Data Group~\cite{pdg06}, $1864.1\pm1.0$~MeV,
we have $-0.9\pm2.1$~MeV for the above difference or for the binding
energy if we assume it is a bound state of $D^0$ and $\bar{D^{*0}}$ mesons.
The large uncertainty in the difference is partially due to the rather
large uncertainty in mass of $D^0$ meson. This was the motivation to
measure $M(D^0)$ more precisely using our $281$~pb$^{-1}$ of data taken
at $\psi(3770)$.

\begin{figure}[h]
\centering
\includegraphics[width=80mm]{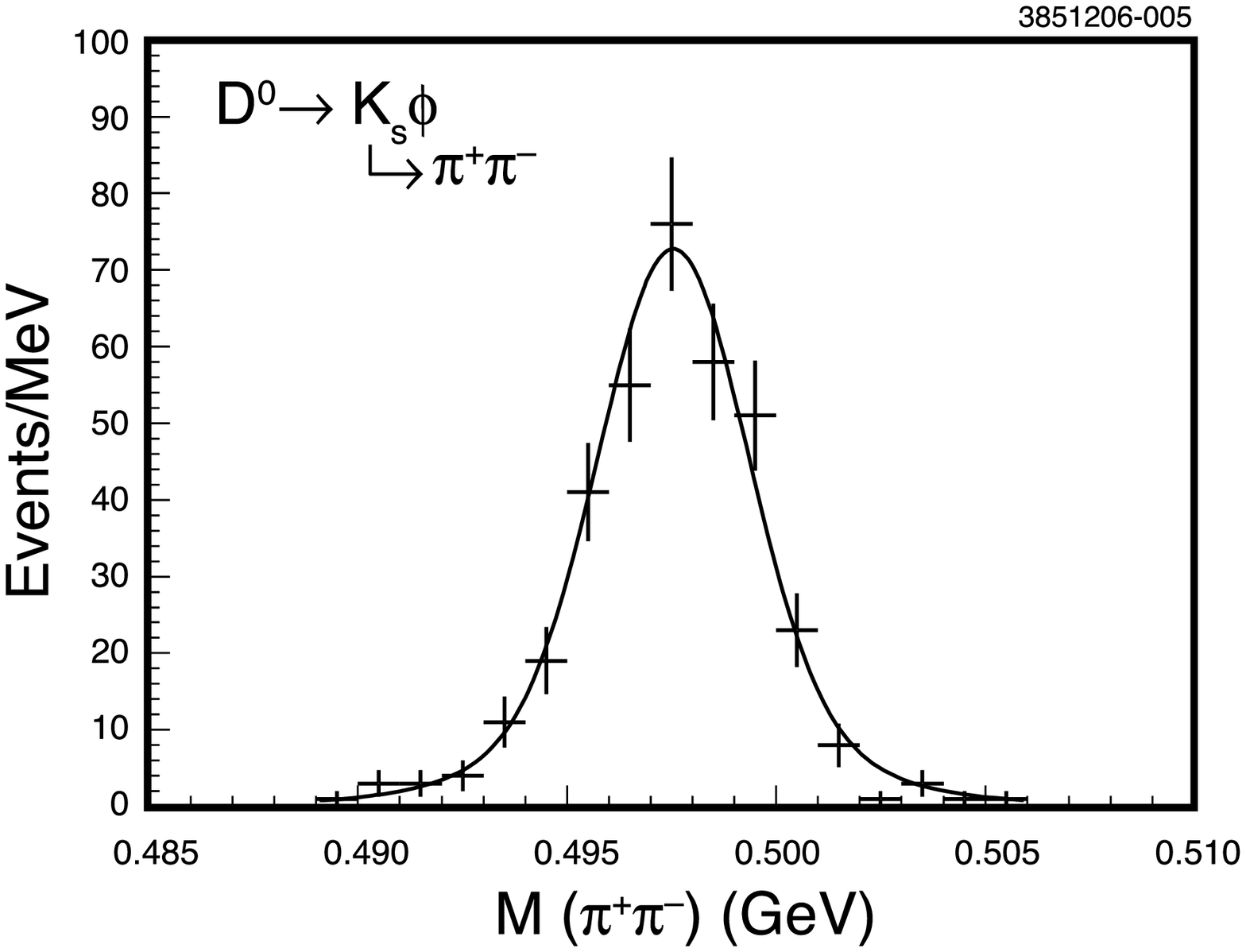}
\caption{Invariant mass of $\pi^+\pi^-$.} \label{fig:xpipi}
\end{figure}

\begin{figure}[h]
\centering
\includegraphics[width=80mm]{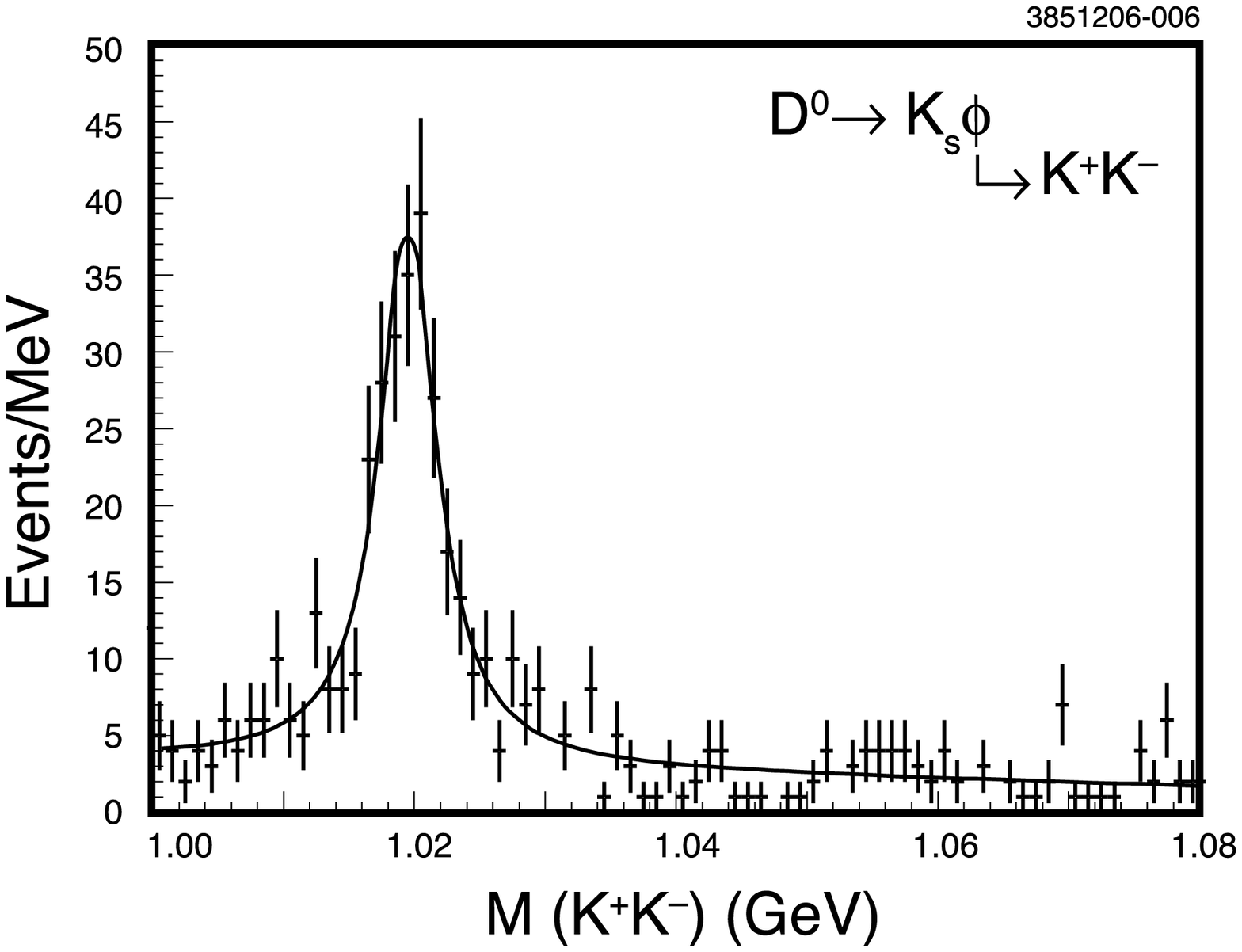}
\caption{Invariant mass of $K^+K^-$.} \label{fig:xkk}
\end{figure}

\begin{figure}[h]
\centering
\includegraphics[width=80mm]{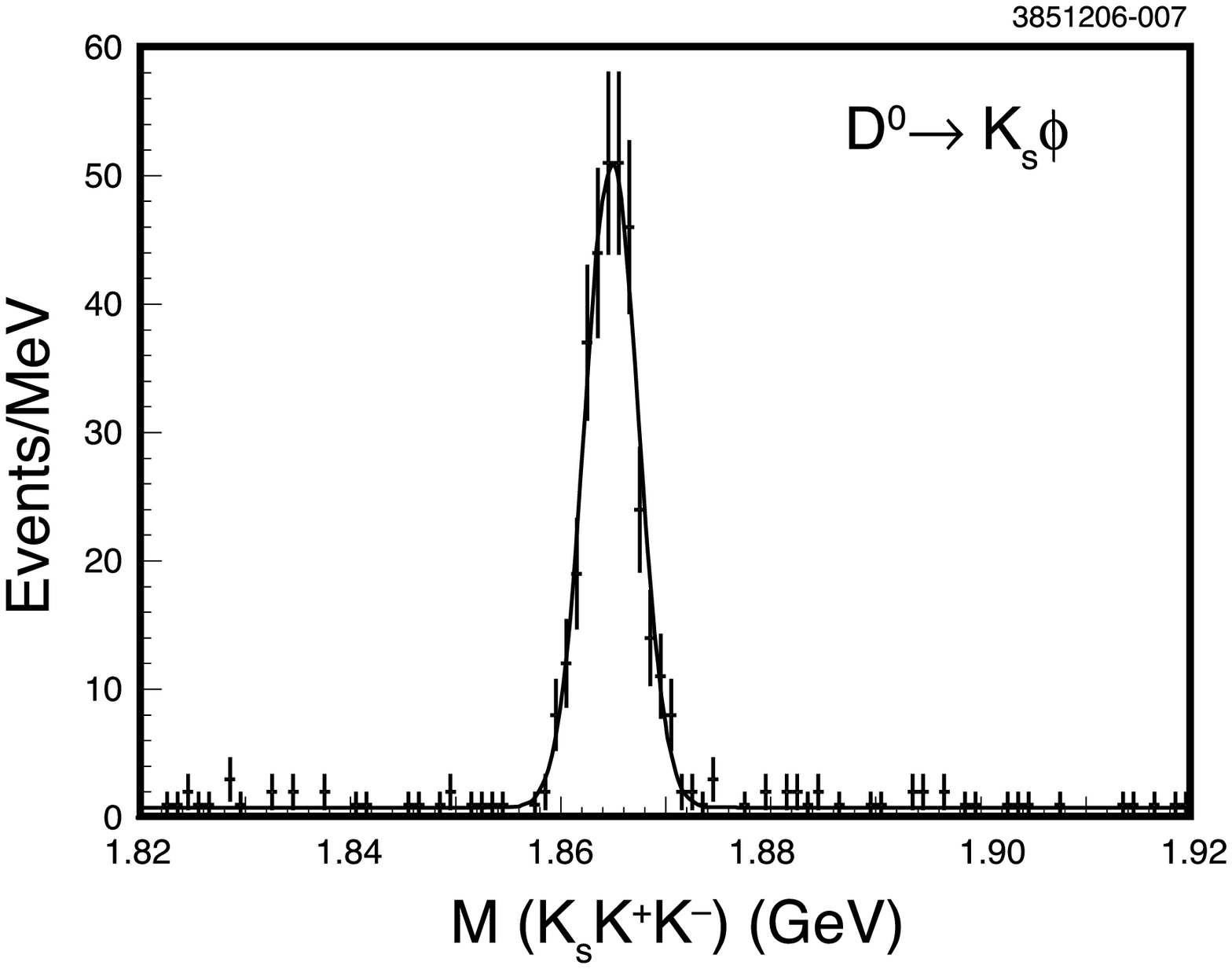}
\caption{Invariant mass of $K_SK^+K^-$.} \label{fig:xd}
\end{figure}

We used a clean (charged particles only) mode of D meson decay,
$D^0\to K_S \phi$ where $K_S\to\pi^+\pi^-$ and $\phi\to K^+K^-$.
Invariant masses of $\pi^+\pi^-$ and $K^+K^-$ are shown in
Figures~\ref{fig:xpipi} and \ref{fig:xkk} respectively.
Figure~\ref{fig:xd} shows the invariant mass of $K_SK^+K^-$ from which
we obtained $M(D^0)=1864.847\pm0.150\pm0.095$~MeV~\cite{xdcleo}.
We then have $M(D^0)$+$M(D^{*0})$-$M(X(3872)) = +0.6\pm0.6$~MeV.
This provides a strong constraint for the theoretical predictions for the 
decays of X(3872) if it is a bound state of
$D^0$ and $\bar{D^{*0}}$ mesons. The uncertainty in its binding energy is
now calling for more precise measurement on mass of X(3872) itself.


\section{Properties of $\eta$}

It has been almost half a century since the $\eta$ meson was
discovered~\cite{etafound}. Since then, many measurements have been
made by many experiments. Still, almost all exclusive branching
fractions are determined as \textit{relatives} to other $\eta$ decays.

Based on the 27M $\psi(2S)$ sample, we measured almost all the major
modes ($99\%$ of generic decays of $\eta$) which allowed us to determine
the major branching fractions~\cite{etarate}.
We obtained the $\eta$ sample through a two-body decay of $\psi(2S)$,
$\psi(2S)\to\eta J/\psi$ where $J/\psi$ subsequently decays to
two leptons ($e^+e^-$ or $\mu^+\mu^-$).
That is, we have sample of about 0.1M $\eta$ decays with a
di-lepton tag on $J/\psi$.

We first constrained the invariant mass of di-leptons to be the known mass
of $J/\psi$. We then combined the fitted $J/\psi$ with $\eta$ decay products
and constrained further to be the mass of $\psi(2S)$. In this analysis,
the $\eta$ decay modes we considered were $\eta\to\gamma\gamma$,
$3\pi^0$, $\pi^+\pi^-\pi^0$, $\pi^+\pi^-\gamma$ and $e^+e^-\gamma$.
According to Ref.~\cite{pdg06}, the sum of these 5 rates amounts to 
$99.88\%$ of the
total $\eta$ decays.
We then took ratios between efficiency-corrected yields separately for each of
$J/\psi\to e^+e^-$ and $\mu^+\mu^-$ cases in which all lepton 
related systematic
uncertainties were canceled. The resulting ratios of $\eta$ branching fractions
are summarized in Table~\ref{tab:etaratios}.

\begin{table}
\caption{Ratios of $\eta$ branching fractions.
For each combination, the efficiency ratio, separately for
$J/\psi \to e^+e^-$ and $J/\psi \to \mu^+\mu^-$, the level of
consistency between the $J/\psi \to e^+e^-$ and $\mu^+\mu^-$
result, expressed in units of Gaussian standard
deviations, $\sigma_{\mu\mu/ee}$, and the combined
result for the branching ratio. 
The dagger symbol indicates that this result is 
most precise measurement to date. 
}
\begin{tabular}{c|c|c|r|c@{ $\pm$ }c@{ $\pm$ }l}
Channel & \multicolumn{2}{c|}{eff. ratio} & $\sigma_{\mu\mu/ee}$& \multicolumn{3}{c}{branching fraction ratio} \\ & $\mu\mu$  & $ee$                &          &\multicolumn{3}{c}{\quad}    \\
 \hline 
$3\pi^0/\gamma\gamma$ & $   0.15$ & $0.15$ & $  1.0$ &  0.884 & 0.022 & 0.019\\ 
$pi^+\pi^-\pi^0/\gamma\gamma$ & $   0.50$ & $0.49$ & $ -2.2$ &  0.587 & 0.011 & 0.009$^\dagger $\\ 
$\pi^+\pi^-\gamma/\gamma\gamma$ & $   0.63$ & $0.60$ & $  0.2$ &  0.103 & 0.004 & 0.004$^\dagger $\\ 
$e^+e^-\gamma/\gamma\gamma$ & $   0.53$ & $0.52$ & $  0.1$ &  0.024 & 0.002 & 0.001$^\dagger $\\ 
$3\pi^0/\pi^+\pi^-\pi^0$ & $   0.30$ & $0.32$ & $  2.1$ &  1.496 & 0.043 & 0.035$^\dagger $\\ 
$\pi^+\pi^-\gamma/\pi^+\pi^-\pi^0$ & $   1.27$ & $1.24$ & $  1.1$ &  0.175 & 0.007 & 0.006 \\ 
$e^+e^-\gamma/\pi^+\pi^-\pi^0$ & $   1.07$ & $1.06$ & $  0.5$ &  0.041 & 0.003 & 0.002$^\dagger $\\ 
$e^+e^-\gamma/\pi^+\pi^-\gamma$ & $   0.84$ & $0.86$ & $  0.0$ &  0.237 & 0.021 & 0.015 \\ 
\end{tabular} 
\label{tab:etaratios}
\end{table}

\begin{figure*}[h]
\centering
\includegraphics[width=135mm]{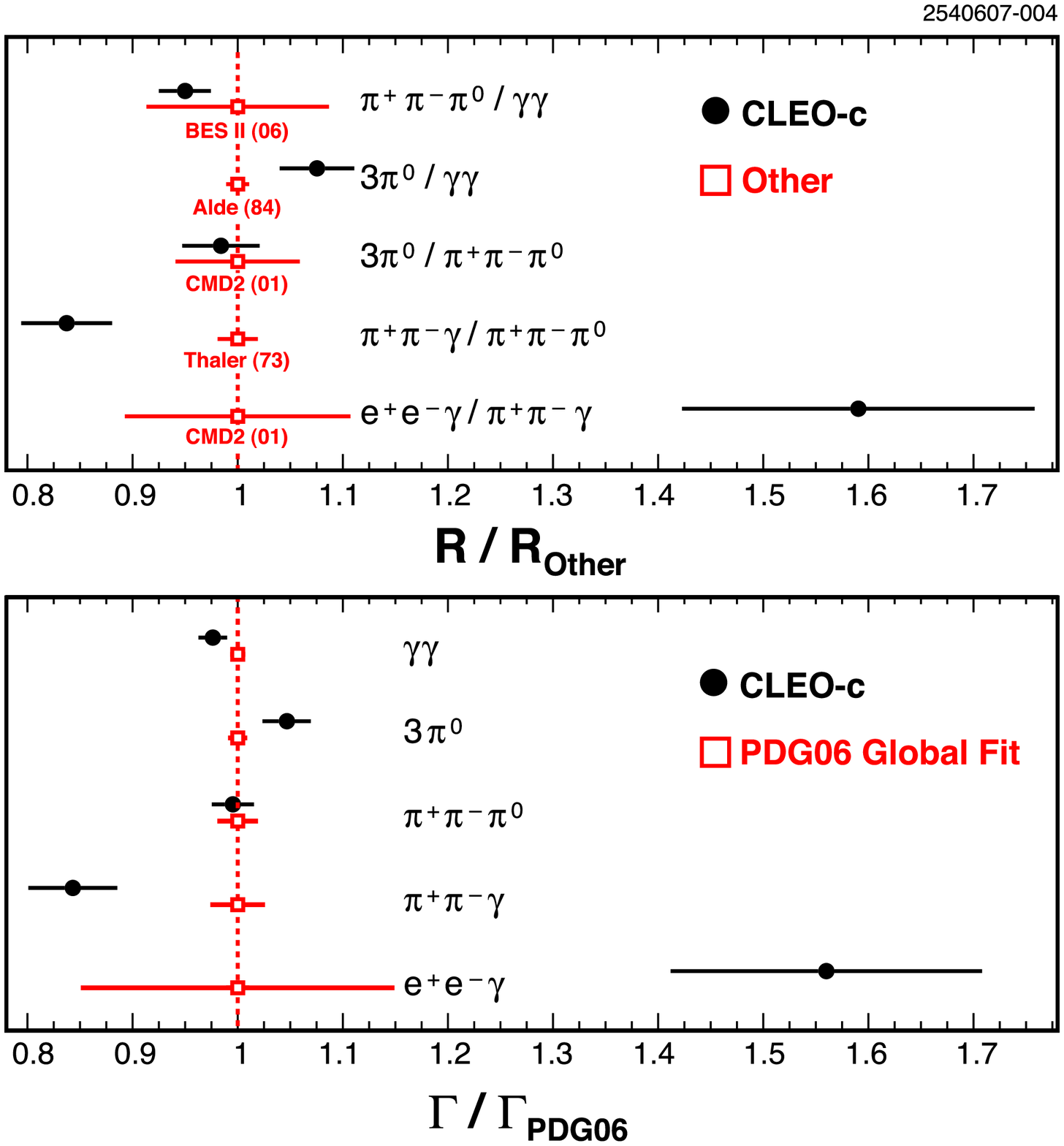}
\caption{Comparison of the results obtained in this analysis with the most
precise measurements from other experiments~\cite{pdg06},\cite{etaother} (top), and the PDG 2006
global fits~\cite{pdg06} (bottom).} \label{fig:etarate}
\end{figure*}

Figure~\ref{fig:etarate} shows a graphical version of comparison in terms of
ratios of branching fractions to the single most precise other measurements
(top of Figure~\ref{fig:etarate}).

By assuming that the 5 exclusive channels we considered in this analysis
cover the all $\eta$ decay modes, we were also able to extract
absolute branching fractions of these 5 modes. Other possible $\eta$ decay
modes are either now allowed and/or found to be less than $0.2\%$ of generic
decays $\eta$~\cite{pdg06}. We included $0.3\%$ as a possible systematic uncertainty
in the absolute branching fraction measurements. The results are summarized in
Table~\ref{tab:etabr} and also shown graphically in Figure~\ref{fig:etarate} in terms of
ratio of our branching fractions to the PDG 2006 global fit
(bottom of Figure~\ref{fig:etarate}). Several of the relative and absolute branching
fractions obtained in this analysis are either the most precise to date or first
measurements.

\begin{table}
\caption{For each $\eta$ decay channel, absolute
branching fraction measurements for $J/\psi \to e^+e^-$
and $J/\psi \to \mu^+\mu^-$ combined, with statistical and
systematic uncertainties (middle column), as determined
in this work. The last column
shows the PDG fit result~\cite{pdg06}.
All but $\gamma\gamma$ are
first measurements.
}
\begin{tabular}{c|r@{ $\pm$ }r@{ $\pm$ }l|r@{ $\pm$ }r}
 Channel & \multicolumn{3}{c|}{this work (\%)}  & \multicolumn{2}{c}{PDG~\cite{pdg06} (\%)} \\ 
 \hline 
$\gamma\gamma$ & $38.45 $ & $ 0.40 $ & $ 0.36$  & 39.38 & 0.26 \\ 
$3\pi^0$ & $34.03 $ & $ 0.56 $ & $ 0.49$  & 32.51 & 0.28 \\ 
$\pi^+\pi^-\pi^0$ & $22.60 $ & $ 0.35 $ & $ 0.29$  & 22.7  \mbox{$\,$}& 0.4\mbox{$\ \,$} \\ 
$\pi^+\pi^-\gamma$ & $3.96 $ & $ 0.14 $ & $ 0.14$  &  4.69 & 0.11 \\ 
$e^+e^-\gamma$ & $0.94 $ & $ 0.07 $ & $ 0.05$  &  0.60 & 0.08 \\ 
\end{tabular} 
\label{tab:etabr}
\end{table}

Further more, we also measured the mass of $\eta$ meson~\cite{etamass}.
This was motivated by two recent precision measurements that
were inconsistent with each other. In 2002, the NA48 Collaboration
reported $M_{\eta}=547.843\pm0.030\pm0.041$~MeV~\cite{na48}, while in 2005,
GEM Collaboration reported $M_{\eta}=547.311\pm0.028\pm0.032$~MeV~\cite{gem}
which was 8 standard  deviations below NA48's result.

We used the same $\eta$ sample described previously in this Section but used
only 4 decay modes, $\eta\to\gamma\gamma$, $3\pi^0$, $\pi^+\pi^-\pi^0$,
and $\pi^+\pi^-\gamma$ while, again, constraining masses of $J/\psi$ and
$\psi(2S)$. Our result, the average of the 4 $\eta$ decay modes, is
$M_{\eta}=547.785\pm0.017\pm0.057$~MeV which has
comparable precision to both NA48 and GEM results, but is consistent with the
former and 6.5 standard deviations larger than the later.
We note that the KLOE Collaboration also recently measured mass of the
$\eta$ meson to be $547.873\pm0.007\pm0.031$~MeV which was presented
at the 2007 Lepton-Photon conference~\cite{kloe}.

\section{Summary}
I have presented confirmation of BaBar's observation of
Y(4260) in di-pion transition to $J/\psi$ along with
a new observation through neutral di-pion transition.
Our precision measurement on $M(D^0)$ calls for more precise
measurement on $M(X(3872))$.
With 3M $\psi(2S)$ sample, we have
results on
two-, three-, and four-body decays of $\chi_{cJ}$ states in which
many sub-structures were seen in three- and four-body modes.
Dalitz plot analyses were done for the case of 3-body decays.
More detailed analyses can be done with the full 27M
$\psi(2S)$ sample. Using the 27M sample, we performed
precision measurements on $\mathcal{B}(\eta\to X)$ and
$M(\eta)$.

\bigskip 

\begin{thebibliography}{99} 

\bibitem{pdg06} W. M. Yao \textit{et al.} [Particle Data Group], J. Phys. G {\bf 33}, 1 (2006).
\bibitem{etaeta} G. S. Adams \textit{et al.} (CLEO Collaboration), Phys. Rev. D {\bf 75}, 071101(R) (2007).
\bibitem{beseta} J. Bai \textit{et al.} (BES Collaboration), Phys. Rev. D {\bf 67}, 032004 (2003).
\bibitem{e823eta} M. Andreotti \textit{et al.} (E-835 Collaboration), Phys. Rev. D {\bf 72}, 112002 (2005).
\bibitem{zhao} Q. Zhao, Phys. Rev. D {\bf 72}, 074001 (2005).
\bibitem{dalitz} S. Athar \textit{et al.} (CLEO Collaboration), Phys. Rev. D {\bf 75}, 032002 (2007).
\bibitem{bes3body} M. Ablikim \textit{et al.} (BES Collaboration), Phys. Rev. D {\bf 74}, 072001 (2006).
\bibitem{y4260ba} B. Aubert \textit{et al.} (\textit{BaBar} Collaboration), Phys. Rev. Lett. {\bf 95}, 142001 (2005).
\bibitem{y4260isr} Q. He \textit{et al.} (CLEO Collaboration), Phys. Rev. D {\bf 74}, 091104 (2006).
\bibitem{y4260sigma} T. E. Coan \textit{et al.} (CLEO Collaboration), Phys. Rev. Lett. {\bf 96}, 162003 (2006).
\bibitem{chirho} X. Liu, X.-Q. Zeng, and X.-Q. Li, Phys. Rev. D {\bf 72}, 054023 (2005).
\bibitem{baryon} C.-F. Qiao, he-ph/0510228 (2005).
\bibitem{x3872bell} S. K. Choi \textit{et al.} (Belle Collaboration), Phys. Rev. Lett. {\bf 91}, 262001 (2003).
\bibitem{x3872cdf} D. Acosta \textit{et al.} (CDF II Collaboration), Phys. Rev. Lett. {\bf 93}, 072001 (2004).
\bibitem{x3872d0} V. M. Abazov \textit{et al.} (D$\emptyset$ Collaboration), Phys. Rev. Lett. {\bf 93}, 162002 (2004).
\bibitem{x3872baba} B. Aubert \textit{et al.} (\textit{BaBar} Collaboration), Phys. Rev. D {\bf 71}, 071103 (2005).
\bibitem{ddbound} E. S. Sanson, Phys. Lett. {\bf B588}, 189 (2004); N. A. T\"ornqvist, Phys. Lett. {\bf B599}, 209 (2004); M. B. Voloshin, Phys. Lett. {\bf B579}, 316 (2004).
\bibitem{xdcleo} C. Cawlfield \textit{et al.} (CLEO Collaboration), Phys. Rev. Lett. {\bf 98}, 092002 (2007).
\bibitem{etafound} A. Pevsner \textit{et al.}, Phys. Rev. Lett. {\bf 7}, 421 (1961).
\bibitem{etarate} A. Lopez \textit{et al.} (CLEO Collaboration), Phys. Rev. Lett. {\bf 99}, 122001 (2007).
\bibitem{etaother} M. Ablikim \textit{et al.} (BES Collaboration), Phys. Rev. D {\bf 73}, 052008 (2006); D. Alde \textit{et al.}, Z. Phys. {\bf C25}, 225 (1984); Yad. Fiz. {\bf 40}, 1447 (1984); R. R. Akhmetshin \textit{et al.} (CMD-2 Collaboration), Phys. Lett. {\bf B509}, 217 (2001); J. J. Thaler \textit{et al.}, Phys. Rev. D {\bf 7}, 2569 (1973).
\bibitem{etamass} D. H. Miller \textit{et al.} (CLEO Collaboration), Phys. Rev. Lett. {\bf 99}, 122002 (2007).
\bibitem{na48} A. Lai \textit{et al.} (NA48 Collaboration), Phys. Lett. {\bf B533}, 196 (2002).
\bibitem{gem} M. Abdel-Bary \textit{et al.} (GEM Collaboration), Phys. Lett. {\bf B619}, 281 (2005).
\bibitem{kloe} F. Ambrosino \textit{et al.} (KLOE Collaboration), arXiv:0707.4616 (contributed paper to Lepton Photon 2007).

\end{thebibliography}

\end{document}